\documentclass[tighten]{aastex63}

\usepackage{xcolor}
\usepackage{ulem,xpatch}

\def\eg{{\it e.g.,} }

\def\ie{{\it i.e.,} }

\def\ha{H{\rm$\alpha$}}
\def\hi{H\,{\sc i}}

\def\he1{He\,{\sc i}}
\def\heii{He\,{\sc ii}}

\def\nii{N\,{\sc ii}}
\def\niii{N\,{\sc iii}}

\def\sii{S\,{\sc ii}}
\def\siii{S\,{\sc iii}}

\def\oiii{O\,{\sc iii}}

\def\ariii{Ar\,{\sc iii}}
\def\ariv{Ar\,{\sc iv}}
\def\cliii{Cl\,{\sc iii}}
\def\cliv{Cl\,{\sc iv}}
\def\mnv{Mn\,{\sc v}}
\def\kms{km\hspace{1pt}s$^{-1}$}
\def\cc{cm$^{-3}$}

\received{?, 2021}
\revised{?, 2022}
\accepted{?}

\submitjournal{ApJ}

\begin{document}

\title{
Double Shells of the Planetary Nebula NGC 7009 Minor Axis}
%Sulfur Forbidden Line Images of the PN NGC 7009 along the Minor Axis}
%[{\sii}] Spectral Images of NGC 7009 along the Minor Axis}

\def\corrauthor{
%%% PUT NAME OF CORRESPONDING AUTHOR HERE %%%%%%%%%%%%%%%%%%%%%%%%%%%
S. Hyung 
%%% END %%%%%%%%%%%%%%%%%%%%%%%%%%%%%%%%%%%%%%%%%%%%%%%%%%%%%%%%%%%%%
}

%%% RUNNING AUTHOR NAME %%%%%%%%%%%%%%%%%%%%%%%%%%%%%%%%%%%%%%%%%%%%%

\def\runningauthor{
%%% PUT RUNNING AUTHOR NAME(S) HERE %%%%%%%%%%%%%%%%%%%%%%%%%%%%%%%%%
%%% END %%%%%%%%%%%%%%%%%%%%%%%%%%%%%%%%%%%%%%%%%%%%%%%%%%%%%%%%%%%%%
}

%%% RUNNING TITLE %%%%%%%%%%%%%%%%%%%%%%%%%%%%%%%%%%%%%%%%%%%%%%%%%%%

\def\runningtitle{
%%% PUT RUNNING TITLE HERE %%%%%%%%%%%%%%%%%%%%%%%%%%%%%%%%%%%%%%%%%%
[Double Shells of NGC7009 Minor Axis 
%%% END %%%%%%%%%%%%%%%%%%%%%%%%%%%%%%%%%%%%%%%%%%%%%%%%%%%%%%%%%%%%%
}

\correspondingauthor{Siek Hyung}
\email{hyung@chungbuk.ac.kr}

\author{Seong-Jae Lee}
\author{Siek Hyung}
\affiliation{Dept. of Earth Science (Astronomy),
  Chungbuk National University, Chungbuk 28644, S. Korea} 

\author{Masaaki Otsuka}
\affiliation{Okayama Observatory, Kyoto University, Kamogata, Asakuchi, Okayama, 719-0232, Japan}

\begin{abstract}
%%% PUT ABSTRACT HERE %%%%%%%%%%%%%%%%%%%%%%%%%%%%%%%%%%%%%%%%%%%%%%%
We analyzed the minor axis spectra of the elliptical planetary nebula (PN) NGC 7009 observed with the Keck HIRES with a 0.862$''$ $\times$ 10$''$ slit placed at about $\sim$7.5$''$ and 10$''$ away from the center and a 0.862$''$ $\times$ 14$''$ slit at the center.
The mean densities derived from the integrated 
[{\sii}] 6716/6731\AA\, fluxes along the Keck HIRES slit length indicate a density range of 10$^{3.7}$ to 10$^{4.1}$ {\cc}, while  the
local densities derived from the slit spectral images show a large local density variation of about 10$^{2.8}$ -- 10$^{4.6}$ {\cc}: 
local densities vary substantially more than values integrated over the line of sight. 
The expansion rates of the {main} and outer shells obtained by [{\sii}] are about 21.7 and  30.0 {\kms}, respectively. 
The kinematic results of the [{\sii}] spectral lines 
correspond to the outermost regions of the two shells and are not representative of the whole PN but are closely related to the other emission lines observed in the shell gas. 
We conclude that the density contrast leads to the formation of the inner shell, while the change in ionization state leads to the formation of the outer shell. 
We suggest that the inner {main} and outer shells result from two successive major ejections. The physical conditions of the central star must have been different when these shells first formed. 
%{\color{blue} The small-sized blobs appear to be accelerated debris leaving the {main} shell due to the strong stellar winds from the central star.   }

%%% END %%%%%%%%%%%%%%%%%%%%%%%%%%%%%%%%%%%%%%%%%%%%%%%%%%%%%%%%%%%%%
\end{abstract}


%%% ABSTRACT %%%%%%%%%%%%%%%%%%%%%%%%%%%%%%%%%%%%%%%%%%%%%%%%%%% 
\section{Introduction}\label{sec:intro} 

Planetary nebulae (PNe) evolved from stars of low-intermediate mass of 0.8 -- 8\,M$_\odot$ reach the final stage of their life after experiencing mass loss during  thermal pulses at their asymptotic giant branch stage. 
During the short life span of thousands of years, the central star of the PN (CSPN) emits UV photons, which ionize the surrounding shell gas, expanding at a velocity of 20 -- 30 \kms. 

The interacting-winds between the hot CSPN and gas shells may have played an important role in the formation of planetary nebulae \citep{1978kwok}. 
The ionized shell gas produces various emission lines, such as the recombination lines of H and He, and the forbidden lines of heavy elements such as N, O, etc.
With high dispersion spectra, one can obtain information regarding the kinematics and geometrical structure of the PN shell and derive chemical abundances.

NGC 7009 is a PN at high galactic latitude ($b$ = $-$34.5$^{\circ}$), whose interstellar extinction at H$\beta$ is very low  $C$ = $\log\,I({\rm H}{\beta})/F({\rm H}{\beta})$ = 0.08 dex \citep{ha95a}, where $I({\rm H}{\beta})$ and $F({\rm H}{\beta})$ are the intrinsic and observed H$\beta$ fluxes, respectively.
The excitation class of the spectrum obtained at the end of the major axis is 5.5, defined by the ratio of [{\oiii}]/{\heii} to [{\oiii}]/{H$\beta$} \citep{1956Aller}. The effective temperature ($T_{\rm eff}$) of the CSPN is about 82\,000~K \citep{kb94}.
Its distance is known to be around 1.2 kpc (1.2 kpc in \citealp{sab04,ode63}, 2 kpc in \citep{kb94}, and 1.15 kpc in \citealp{2018kim}: GAIA).

The derived abundances with the help of photo-ionization (P-I) codes such as NEBULA by \citet{hyu94} or { CLOUDY by \cite{2017RM} show } that NGC 7009 is an O-rich object with an O/C $> 1$ \citep{ha95a, ha95b}.
Although the derived abundances indicate NGC 7009 to be a classic Galactic PN, specific lines such as [{\nii}] and {\heii} show unusually strong intensities that could not be accommodated by a typical ionization stratification effect  \citep{ha95b, Gon03}.

NGC 7009 is morphologically classified as an early middle or middle elliptical PN, with
a fast isotropic stellar wind in the inner part, surrounded by an inner halo inside the  main shell \citep{1987B2BPV}. 
The compact and  bright nebula is suitable for investigating the details of substructures related to its kinematics or evolutionary status. The high dispersion
long-slit spectrophotometric observational data of the bright rims, knots, and halos will allow us to define better the kinematic structures of density and temperature, involving the ionization or possible shock in the shells and the outer halo zone  (\citealt{1985RA, 1987B1, 1987B2BPV, 1997Mellema}).
%\citet{lame1996} derived the electron density map using the two Fabry-Pero interferometry, narrow filter [{\sii}] 6716/6731\AA\,  images for the whole NGC 7009 nebula.  
% Their electron density map in the central region indicates that  $N_{\rm e}$ $>$  8000 {\cc}, showing an elliptical high-density region with its  major axis rotated clockwise 50$^{\circ}$ relative to the major axis of NGC 7009.
\citet{lame1996} derived the electron density map for the whole NGC 7009 nebula, which indicates $N_{\rm e}$ $>$  8000 {\cc} in the central region.

Abundance mismatches derived from recombination and forbidden lines present in some PNe were hypothesized to arise partly due to temperature fluctuations \citep{Pei67}.
\citet{rub02} conducted Hubble Space Telescope ($HST$) WF/PC2 imaging and STIS long-slit spectral studies for NGC 7009 and they found a rather uniform $T_{\rm e}$ map within the ranges of 9000 -- 11\,000~K across the nebula with a very low fractional mean-square $t^{2}$ defined by \citet{Pei67}. 

Recently, \citet{2016walsh,2018Walsh} performed similar but more advanced studies with the MUSE integral field spectrograph on the ESO VLT. From various available emission lines from neutral to the highest ionization ({\heii} and [{\mnv}]), they presented the entire surface maps of electron temperature (and $t^{2}$) and density, which are averaged across the line { of sight. }
Using collisional de-excitations of the { [{\sii}] $^2 D_{3/2,5/2}$ levels, } they derived density maps showing a wide density range of $N_{\rm e}$ $\sim$ 3000 -- 8000 {\cc} in the inner bright zone.

Long-slit observations provide information for large spaces from a small number of lines with a limited wavelength range. 
Echelle spectral observations, such as the Hamilton Echelle Spectrograph (HES) at Lick Observatory,  have a high dispersion ability to separate  closely-spaced lines for many lines.
However, the HES provides no spatial information.
The Keck HIgh-Resolution Echelle Spectrograph (HIRES) has the advantage of long slit spatial dispersion and high echelle  dispersion capabilities.
We analyze the two-dimensional (2-D) [{\sii}] 6716.4 and 6730.8\AA\, spectral images secured with the Keck HIRES along the minor axis of NGC 7009.

The collisional excitation of singly ionized sulfur, or [{\sii}],
has relevance to the study of {the Sun,  the Jupiter-Io in the solar system \citep{1994Io}, symbiotic stars \citep{1990Sch}, nebulae \citep{1992Ost}, and active galactic nuclei \citep{1999Ost}. }

Keenan et al. (1996, K96) investigated the [{\sii}] transitions  to the ground state from 
two separate metastable states, namely via the de-excitation of the above-mentioned (1)
$^2$D  $\rightarrow$ $^4$S  (6716\AA\, and 6731\AA) and (2) $^2$P  $\rightarrow$ $^4$S  (4069\AA\, and 4076\AA) transitions. They showed that both  are sensitive to both electron  temperatures and densities up to  log~$N_e$ = {5} dex {\cc}. 
When calculating the line strengths of [\sii] 6716\AA\, and 6731\AA,
we used { the 2017 release of Cloudy code (sub-version 17.02)   \citep{2017RM}.\footnote{ https://gitlab.nublado.org/cloudy/cloudy/-/wikis/NewC17} }

Section~\ref{sec:obs} describes observations of the   position-velocity (P-V)   slit spectral images by Keck HIRES and the data reduction. In Section~\ref{sec:p-v}, we derive position velocity maps of the density distribution for the { innermost bright main} and outer shells along the minor axis, respectively. We
update the radial variation of the expansion velocities of the two shells.
In Section~\ref{sec:con}, we give conclusions.

\section{Observations and Data Reduction}\label{sec:obs}

The minor axis of NGC 7009 was observed with the Keck HIRES on August 14th -- 15th, 1998 (UT) by Hyung and Aller. The {seeing was } slightly less than 1.0$''$ in observation. 
Several different HIRES slit settings were used because the bright nebular rim size is about 28$''$ $\times$ 30$''$ and the slit length of the HIRES is smaller than the nebular rim.

\begin{figure*}
\centering
\includegraphics[width=0.6\textwidth]{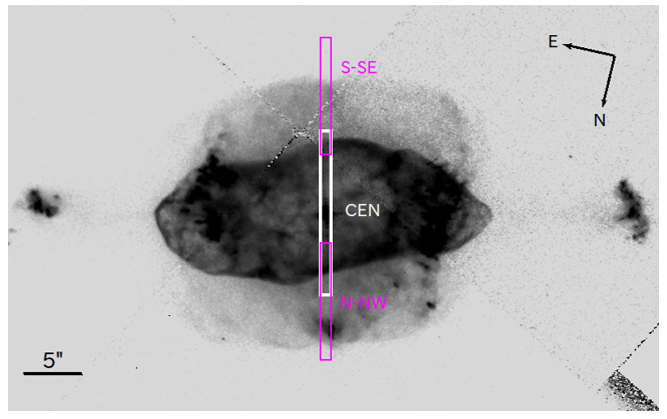}
\caption{HST image of NGC 7009 and Keck HIRES slit direction on the nebular image. The HST WFPC2 image with the F658N filter shows main shell, outer shell, and caps/ansae  (NASA/ESO).  The 10$''$ or 14$'' \times$ 0.862$''$ slit entrance was placed at the S-SE, the center (Cen),  and the N-NW areas along the minor axis  (PA $= +347^{\circ}$).   South (S) is upward, and North (N) is downward. See the text.
 }
\label{fig1}
\end{figure*}

The HIRES has a separation of 8 -- 43$''$ between the two adjacent echelle orders, which enables us to obtain spatial information along the slit length direction.
Table~\ref{tbl-1} lists the employed slit information.
The [{\sii}] spectral images  were secured with two slit (length $\times$ width) sets:  (a) s1: 14$''$ $\times$ 0.862$''$ (6309 -- 8530\AA); and  (b) s2: 10$''$ $\times$ 0.862$''$ (5130 -- 7475\AA).
The slit {width corresponds } to a wavelength dispersion power of $R = \lambda /\triangle \lambda$ = 45\,200 ($\sim$3 pixels).

\begin{table*}
\centering
\caption{Keck HIRES observation.
 Two slit lengths, 14$''$ (s1) and 10$''$ (s2), with the same slit width were adopted. Column (5) indicates the distance of the slit center from the central position.  See Figure~\ref{fig1} for the position angles of the slit length direction adopted along the  minor axis (PA = $+347^{\circ}$).
}
%\vspace{-0.5cm}
\begin{tabular}{@{}lccccc@{}}
\hline
\hline
{Slit center} & {Date } & {Exp. } & {Slit setup } & Slit center  &   \\
  position &    (UT) &  (min)  & (length) & distance &   \\
\hline
center  & 1998.08.15 & 2  & { s1 (14$''$)  } & &   \\
S-SE    & 1998.08.14 & 25 & { s2 (10$''$) } &  $\sim$10$''$ &  \\
               & 1998.08.14 & 3 & { s2 }  & &   \\
%               & 1998.08.16 & 15 & s3 (7$''$) & .. \\
%               & 1998.08.16 & 2 & s4 (3.5$''$) & .. \\
N-NW   & 1998.08.14 & 30 & { s2 } & $\sim$7.5$''$  &   \\
               & 1998.08.14 & 3 & { s2 } &  &   \\
%               & 1998.08.16 & 20 & s3 &  {\color{blue} $+347^{\circ}$ }\\
%               & 1998.08.16& 2  & s4 & .. \\
%               & 1998.08.16 & 5  & s4 & .. \\
\hline

\end{tabular}
\label{tbl-1}
\end{table*}

Figure~\ref{fig1} shows the nebular image from the Hubble Space Telescope (HST) WFPC2  with the F658N filter (PI: Balick, Prop-ID: 6117, NASA/ESA), with a schematic diagram  of the employed HIRES slit directions and positions over the nebular image:  the  S-SE, Center, and N-NW {pointings} respectively, along the minor axis (PA $= +347^{\circ}$ or $-13^{\circ}$).  Here, we used a non-standard PA value to indicate the current image is an upside-down version of the { standard $167^{\circ}$ case.}

The 2-min exposure with the 14$''$ slit was done at the  CSPN position, while 3-, 25-min, and 30-min exposures with the 10$''$ slit were 
carried out to cover two bright rims along the minor axis. Here, the locations of the S-SE and N-NW slit centers were at $10''$ and  $\sim7.5''$ from the CSPN, respectively.
A standard star 58 Aql suitable for the red wavelength region was observed for flux calibration.

We used a 2048 $\times$ 4096 pixels CCD in the observation, whose 2048 pixels correspond to the spatial direction (\ie the slit length).
The observed spectral data were reduced with the {\sc IRAF}\footnote{The  NOAO Image Reduction and Analysis Facilities {\sc IRAF}
  is distributed by the National Optical Astronomy Observatories,
  operated by the Association of Universities for Research in
  Astronomy (AURA), Inc., under a cooperative agreement with the
  National Science Foundation.} reduction package. 
The {\sc IRAF} long-slit standard data reduction procedure, `noao.twodspec,' was employed to find the spectral image, keeping the spatial information: (a) zero-intensity level correction to all frames including bias, flat lamp, object, and Th-Ar comparison frames using the overscan region of each frame and the mean bias frames; (b) cosmic ray removal; (c) flat-fielding; (d) a 2-D position-velocity spectrum extraction; (e) wavelength calibration keeping the spatial information; and (f) flux calibration using with 58 Aql.
We derived a (heliocentric) radial velocity  of $-$48.91 {\kms}, from the strong lines ({\it i.e.}, F($\lambda$) $>$ 1.0 on the scale of F(H$\beta$) = 100) from the {\sc IRAF} 1-D standard data reduction procedure, 'noao.echelle'  for the same Keck HIRES data. 
Our value is close to the radial velocity of $-$49.0 {\kms} derived through the HES data at Lick Observatory   \citep{ha95a}. 
The derived radial velocity would allow us to convert the observed wavelengths  ($\lambda_{\rm o}$) of two spectra to the ones at the proper coordinate system  of the PN, 
$\lambda_{\rm p}$ = $\lambda_{\rm o}$ + $\Delta\lambda$ where $\Delta\lambda$ = $+$(48.91/c)$\times$ $\lambda_{\rm th}$ and $\lambda_{\rm th}$ is the theoretical wavelength (6716.47\AA\, and 6730.85\AA: \citealt{1972M, 1993M}). The redefined wavelengths would become the kinematic reference point.

Several painstaking steps are required to obtain the electron density from the wavelength calibrated flux via IRAF (see  Figures~\ref{fig4} and \ref{fig5}).   We converted the wavelength to {\kms} for the x-axis to avoid possible errors during the reduction process (\eg  the superposition or interpolation of a [{\sii}] 6716 line profile and another [{\sii}] 6731 one).   Then, we expect the observed line centers of the two theoretical [{\sii}] 6716.47\AA\, and [{\sii}] 6730.85\AA\, lines to correspond V$_{\rm s}$ = 0 {\kms}.

After obtaining the   continuum subtracted flux values on a 2-D position-wavelength plane with the {\sc IRAF}, we further analyze them on a 2-D position-velocity (P-V) plane with reduction algorithms written in  Starlink/Dipso\footnote{http://starlink.eao.hawaii.edu/starlink} and Interactive Data Language (IDL)\footnote{http://idlsgroup.com}. 
Finally, we derive density distribution P-V maps from the   [{\sii}] 6716,6731 PV diagrams.

\section{The P-V Density Distribution Map} \label{sec:p-v}

The 3-D structure models by \citet{sab04} and \citet{ste09} showed that the {main} shell of NGC 7009  is a density contrast prolate ellipsoidal shell. 
The 2-D {\hi} or {\ha} spectral images will display the characteristics of the overall kinematics of the PN. Meanwhile, 
the 2-D [{\sii}] spectral images would give us information of the [{\sii}] {subshell} that {\hi} does not reveal:  a {\hi} line broadening is about six times larger than a  [{\sii}] line for the same electron temperature.

\subsection{Two Main Shells} \label{sec:two}

\subsubsection{The H{$\alpha$} P-V map} \label{sec:ha}

\citet{sab04}  presented the  [{\nii}], {\ha},  [{\oiii}], and {\heii} narrow filter slit images for 12 PAs and constructed 
a three-dimensional structure  with the help of photo-ionization  models.
In addition, they presented the radial variation of ionization structures (He$^+$, He$^{++}$, O$^{\rm o}$, O$^+$, N$^+$, Ne, S$^+$, S$^{++}$, Ar$^{+++}$, Ar$^{3+}$, relative to O$^{++}$) across the minor axis (PA = 169$^{\circ}$) and the major axis (PA =  77$^{\circ}$). The ionization regions of S$^+$ and N$^+$ extend farther to the N-NW than to the   S-SE  direction. Unlike the other six ions, the intensities of  S$^+$ and N$^+$ relative to H$^+$are higher in the outer part,
perhaps due to the hardening and dilution of the UV radiation field from the central star in the outer region due to its distance from the central star.

The spectral images by \citet{sab04}     show that the outer [{\nii}] shell is a structure of a  complex, broad shell filled with material, which appears to be detached from the bright main shell, while the  {\hi} and [{\oiii}] images look like a part tied to the inner main shell (see their Figures 2 and 3). Meanwhile, the main shell shows a possible thin triaxial ellipsoid broken along the major axis 
(projected in PA =  77$^{\circ}$).   
This is not a typical case for the equatorial region slit spectra of a prolate shell, implying 
a possibility that the two shells were formed by different mechanisms.  The four [{\nii}], {\hi}, [{\oiii}], and {\heii} PV spectral images for the minor axis by \citet{sab04} show that the velocity structure of the bright inner rim and the matter outside the rim are different. The integrated image in Figure 1 also clearly shows that the physical structure is similar to what they found. 
The radial distribution of the main shell shows a sharp, bell-shaped profile with a 
 $N_{\rm e}$([{\sii}])  density  being up to 4000 ($\pm$500) {\cc} along the major axis and 7000 ($\pm$500) {\cc} along the  minor axis  \citep{sab04}.  
Our present study on HIRES 2-D spectral images refine this earlier extensive work on the minor axis by \citet{sab04}. 
The two velocity ellipses show misalignment.  We try to 
derive expansion velocities and densities more accurately, taking into account the orientation of the shells based on the [{\sii}] spectral images secured along the minor axis.

\begin{figure*}
\centering
\includegraphics[width=0.80\textwidth]{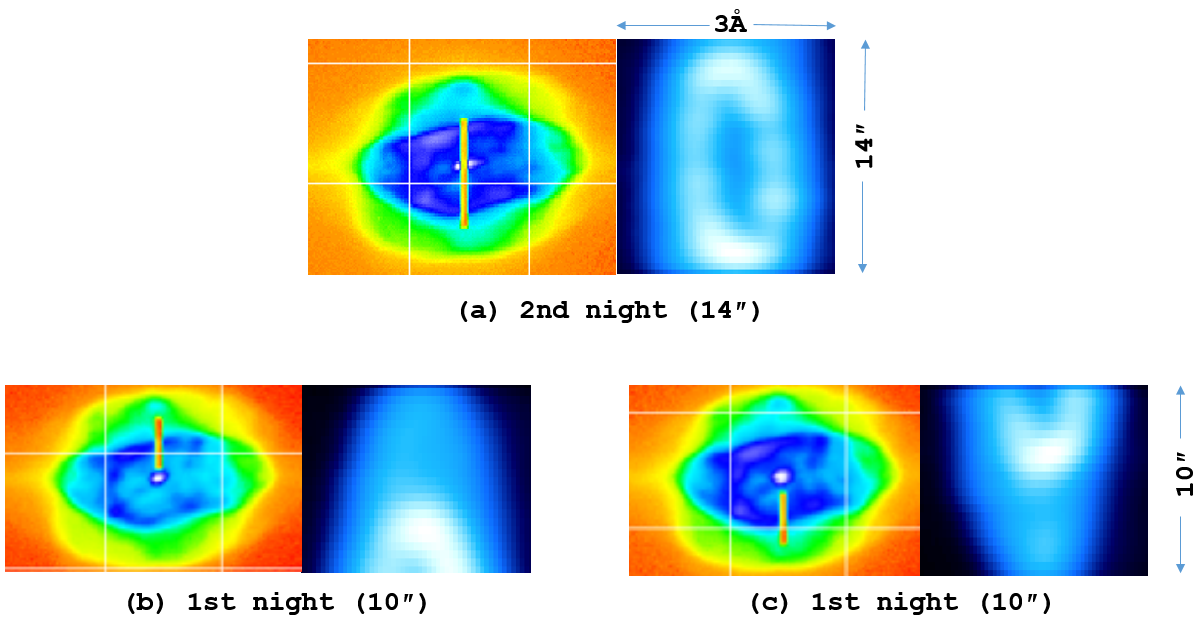}
\caption{The slit monitoring camera images(left) and the reduced 2-D {\ha} spectral images (right).  Obs. date: Aug. 14th, 1998 UT (1st night) and Aug. 15th, 1998 UT (2nd night).  
The spectral image is on a log scale showing that the white color represents an intense flux.
x-axis: full size is 3\AA. The right side of the 2-D spectrum is the redder, \ie receding wavelengths.
    y-axis: 14$''$ or 10$''$ in the sky. (a):  the 14$''$ slit at the CSPN position along the minor axis.  (b) and (c):  the  10$''$ slit at the  S-SE and N-NW directions, respectively. See the text and Table~\ref{tbl-1}. 
 }
\label{fig2}
\end{figure*}

Figures~\ref{fig2}  (a), (b), and (c) show two different images: the green-blue false colors on the left are slit images obtained during observation, while  the white-blue false colors on the right are images of the {\ha} spectrum obtained by analyzing the light entering the slit entrance. 
 The oval 2D shape of the image is artificial in that the x-axis is the wavelength 
component in the units of \AA, and the y-axis is the spatial component in arcsec.
The box size in Figure 2 (a) is
14$''$ $\times$ 3\AA. 
Note that the top is oriented S-SE, and the bottom is N-NW.

The {\ha} spectral image along the minor axis in Figure~\ref{fig2} (a) generally shows a bilaterally symmetric and roughly point-symmetric elliptical rim. 
Still, the  red-side rim appears to be broken, implying an activity of deviating from the toroidal symmetric  
 structure. 
Figuring out the physical condition or identifying  unique substructures could help allow us to understand the history of PN formation or evolution. 
The  {\ha} spectral images help to know  the overall structure of the nebula or where most of the mass is.

The S-SE and N-NW   spectral images in Figure~\ref{fig2}  clearly show the middle zone located outside the bright {main} shell.
\citet{hyu14} performed a somewhat preliminary analysis of Keck spectral images to find the radial velocity variations of the inner bright  shell ({main} shell) and the faint outer shell. Their analysis indicates the Hubble-type expansion of two shells with a 
 velocity jump or a non-zero starting velocity  near the inner shells. We will update these results in this study.   
Figures~\ref{fig2} (b) and (c) show a slight increase in the  {\ha} emission at the extremes of the outer shell. 
The {\ha} width due to the thermal line broadening using the same $T_{\rm th}$ = 10\,000~K would be much broader than that of [{\sii}], \ie 21.6 {\kms} vs. 3.82  {\kms}, so the {\ha} emission line is not suitable for investigating the boundary structure of the shells. Note that the wavelength dispersion width (3 pixels due to the employed slit width) is 6.7 {\kms}.

The appearance of a double-shell structure (or rim) might consist of (1) the inner relatively high-density main shell and (2) the bright outer region of [{\nii}] or [{\sii}] emission by increasing the fractional ionization in this region due to the diluted UV radiation. As a result, the double shells shape might appear. However, looking at the radial variation of logarithmic [{\oiii}] flux along the minor axis (PA=169$^{\circ}$) in Figure 7 by \citet{sab04}, we can see the continuous flux distribution between the two shells. Still, two flux peaks appear about at the same position as the other low excitation lines, and the double shells are also well defined in this medium excitation [{\oiii}] line image: 
The relative peak fluxes of the main shell, intermediate zone, and outer shell are approximately 1, 0.5, and 0.8, respectively. 

The multiple shells discussed in this study or \citet{sab04} could be part of a single structure, as the surface brightness profile, \eg the $HST$, shows an ellipsoid-like structure. The mechanism of the main shell and the outer shell having different shapes will require an ad hoc modification of the radial variation of the densities and physical condition for fractional ionization to produce the observed double shell feature.

\begin{figure*}
\centering
\includegraphics[width=0.60\textwidth]{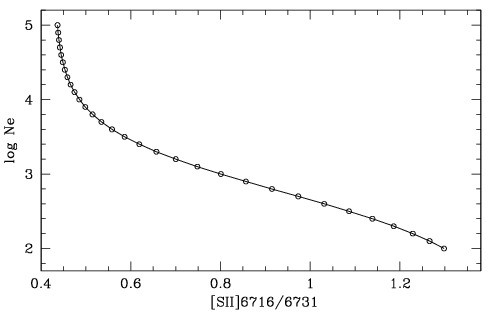}
\caption{{ Density vs. [{\sii}] 6716/6731\AA\ ratio. The P-I model by Ferland et al. (2017) is indicated with open circles, assuming $T_{\rm e}$ = 10\,000 K. See the text. }
 }
\label{fig3c}
\end{figure*}

\subsubsection{Densities averaged over the line of sight} \label{sec:sight}

\citet{2018Walsh} derived the density maps over the whole nebula image, corresponding to the 
integrated fractional volume across the line of sight, based on the [{\sii}] 6716/6731\AA\, and [{\cliii}]5518/5538\AA\, line ratios.
The derived former density map showed a wide density range of $N_{\rm e}$ $\sim$ 3000 -- 8000 {\cc} in the inner bright zone,
especially the high-density in the bright rims (the {main} shell) of the minor axis. Meanwhile, the latter map indicates a lower density range of $N_{\rm e}$ $\sim$ 3000 -- 5500 {\cc} (see their Figure 7). 
The Paschen lines are in a difficult part of the spectrum to measure carefully due to 
atmospheric emission and absorption. 
The densities derived by the hydrogen {\hi} Paschen lines (P15 -- P26) also suggest a presence of much higher densities $\sim 15\,000$ -- 20\,000 {\cc}  in the minor axis zone (see their Figure 11).
These density maps, however, represent the average value of the pencil-beam cross-section and the actual density consists of higher and lower density ones. 
Their [{\sii}], [{\cliii}], and {\hi} density maps generally show similar morphology and structures, but amplitudes and absolute 
values vary depending on the ion used, perhaps due to atomic data.

In the  narrow-band filter [{\sii}] 6716/6731\AA\ Fabry-Perot interferometry study, 
\citet{lame1996} identified an elliptical high-density structure with a maximum density of 10\,000 {\cc} with its major axis deviating  50$^{\circ}$ from the major axis of the NGC 7009 rim. 
Their results mostly agree with \citet{2018Walsh}. 
The densities of the minor axis appear to be higher than those of the major axis. 

In the analysis of NGC 7009 major-axis HST data, \citet{phi10} found the [{\sii}] 6716/6731 ratios of regions corresponding to  0 -- $5''$ and 5  -- $15''$ to be 0.59 -- 0.72 and 0.5 -- 0.65, respectively, corresponding to density ranges of $N_{\rm e}$ $\sim$ 10$^{3.2}$ -- 10$^{3.5}$ and 10$^{3.5}$ -- 10$^{3.8}$. 
In the outer 10 -- $25''$ region, the density ranges are comprehensive, $N_{\rm e}$ $\sim$ 10$^{2.7}$ -- $>$10$^{5}$  corresponding to ratios of 0.37 - 0.95. 
\citet{ha95b} gave the density values (log~$N_{\rm e}$ = 3.55 -- 4.3 {\cc}) derived from the Lick HES lines for the minor axis bright rim  at 5 -- 6$''$ from the center assuming $T_{\rm e}$ = 10\,000 K  (see their Figure 1).

\subsubsection{Density from 2-D [S\,{\sc ii}] maps}

Figure~\ref{fig3c} shows the density diagnostic diagram of electron densities vs. [{\sii}] 6716/6731\AA\ ratios,  produced by Cloudy P-I calculations assuming $T_{\rm e}$ = 10\,000 K with Cloudy 17.02 (\citealt{2005Ir, 2010Tayal}). For spectral image analysis, we used this diagnostic diagram. 

\begin{figure}
\centering
\includegraphics[angle=0,origin=c,width=80mm]{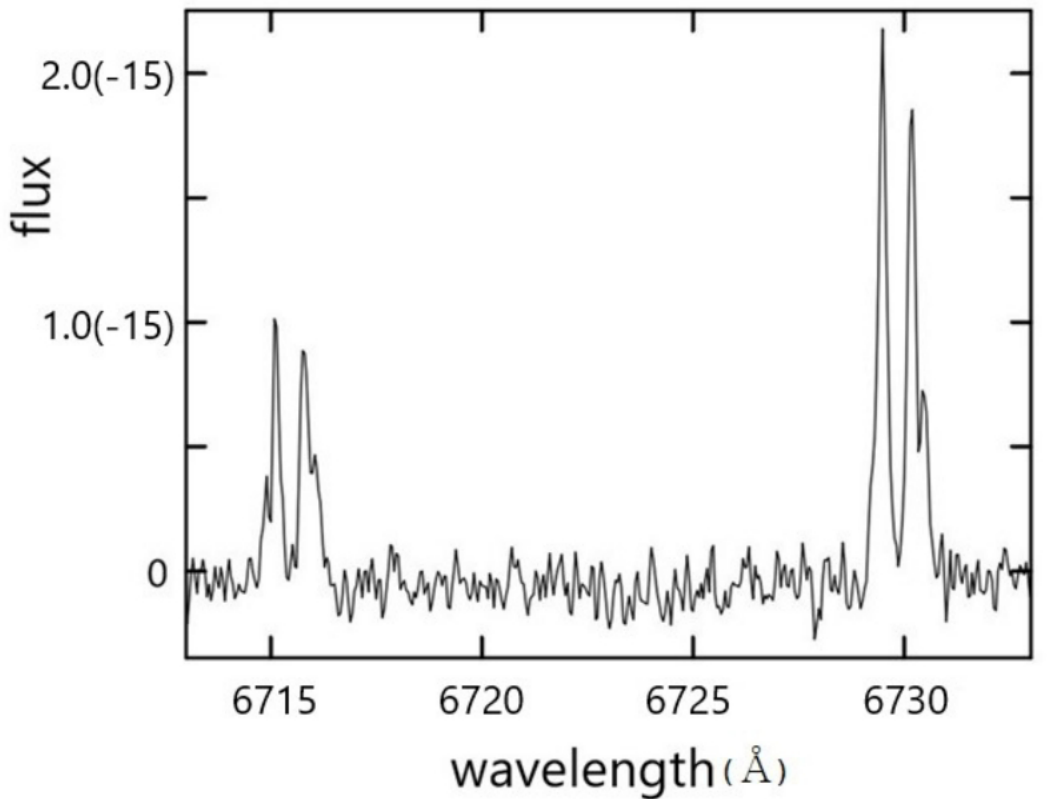}
\vspace{0pt}
\caption{
The spectral scan, showing the [{\sii}] 6716\AA\, and [{\sii}] 6731\AA\, lines in one echelle order. The flux  on the y-axis is in units of erg$^{-1}$ $\AA^{-1}$ $cm^{-2}$ s$^{-1}$ (\eg 2.0($-$15) stands for 2.0 $\times 10^{-15}$). The spectral scan was from { one spatial component or a single row of the CCD,} corresponding to the 4$''$ N-NW position from the center (see Figure~\ref{fig2} (a)). Each pixel corresponds to $\sim$0.39$'' \times \triangle\lambda$ where $\triangle\lambda$ =  $\lambda$/(3$R$).  The continuum level has been adjusted. See the text. 
 }
\label{fig4}
\end{figure}

Figure~\ref{fig4} shows  the [{\sii}] 6716 and [{\sii}] 6731 lines in one echelle order. 
We checked the line ratio from the integrated [{\sii}] line fluxes at a specific spatial position from the Keck HIRES data as in Figure~\ref{fig4}, which shows that the ratio does not differ from the Lick HES [{\sii}] line ratio.

\begin{figure}
\centering
\includegraphics[width=80mm]{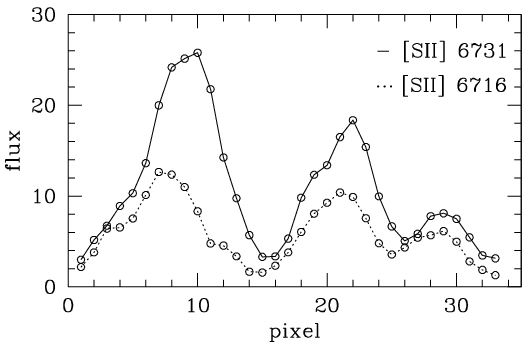}
\vspace{-5pt}
\caption{ [{\sii}] 6716\AA\, vs. [{\sii}] 6731\AA\, flux intensities along a fixed spatial component, \ie 4$^{\prime\prime}$ sky (see Figure~\ref{fig6} (e)).  [{\sii}] 6716\AA\, (dotted) and [{\sii}] 6731\AA\, (solid) flux intensities (10$^{-16} $erg$^{-1}$cm$^{-2}$s$^{-1}$pixel$^{-1}$) with both continuum level and interstellar extinction corrected.  The x-axis is the pixel corresponding to the wavelength.  See the text and Table~\ref{tbl-2}.
}
\label{fig5}
\end{figure}

Figure~\ref{fig5} shows the two flux profiles of [{\sii}] 6716\AA\, and  [{\sii}] 6731\AA. 
These two flux profiles  are replotted to overlap with each other with respect to pixel numbers in Figure~\ref{fig5}.
To estimate fluxes for [{\sii}] 6716\AA\, and  [{\sii}] 6731\AA\, at each pixel, we have to remove the continuum flux.  
After determining the continuum level with a first-order spline function in the raw spectral scan avoiding the emission [{\sii}] line spectrum itself, we subtract the continuum from the raw line spectral scans  (Figure~\ref{fig4}).
As in Figure~\ref{fig4}, the spectral noise in one spectral scan is more or less constant in all wavelength ranges, while the flux level of [{\sii}] varies according to the shape of the line profile. The S/N value becomes worse at a weak flux point.

We also corrected for interstellar extinction with the extinction coefficient of $C = 0.08$.    
Two clearly well-defined rims of the main shell and the other outer shell are seen in both  lines. 
Figure~\ref{fig5} shows two overlapping spectra extracted from a fixed spatial component as in Figure~\ref{fig4}.

The x-axis in Figure~\ref{fig5} is the pixel corresponding to the wavelength. 
One must correct the wavelength dispersion first to calculate the line intensity ratio.  
However, plotting Figure~\ref{fig5} as flux versus velocity would simplify the necessary task.

The obtained P-V density map should match the others in the overlapping zone. The wavelength identification was done with the {\it IRAF} 2D long-slit data reduction procedure (noao.twodspec) and the whole 2D spectral image had to be aligned.  As with all other echelle orders, the HIRES spectral scans are  also slightly curved when recorded on a CCD of the spectrometer. Hence, one needs to do an IRAF reduction procedure that rotates the 2D raw image to align the 2D image parallel to the horizontal axis. These processes may lead to errors in wavelength identification. However, the error is not significant for our [{\sii}] spectral scan echelle order.
Note that there is an overlapping zone between the two different slit spectral images.

\begin{figure*}
\centering
\includegraphics[width=160mm]{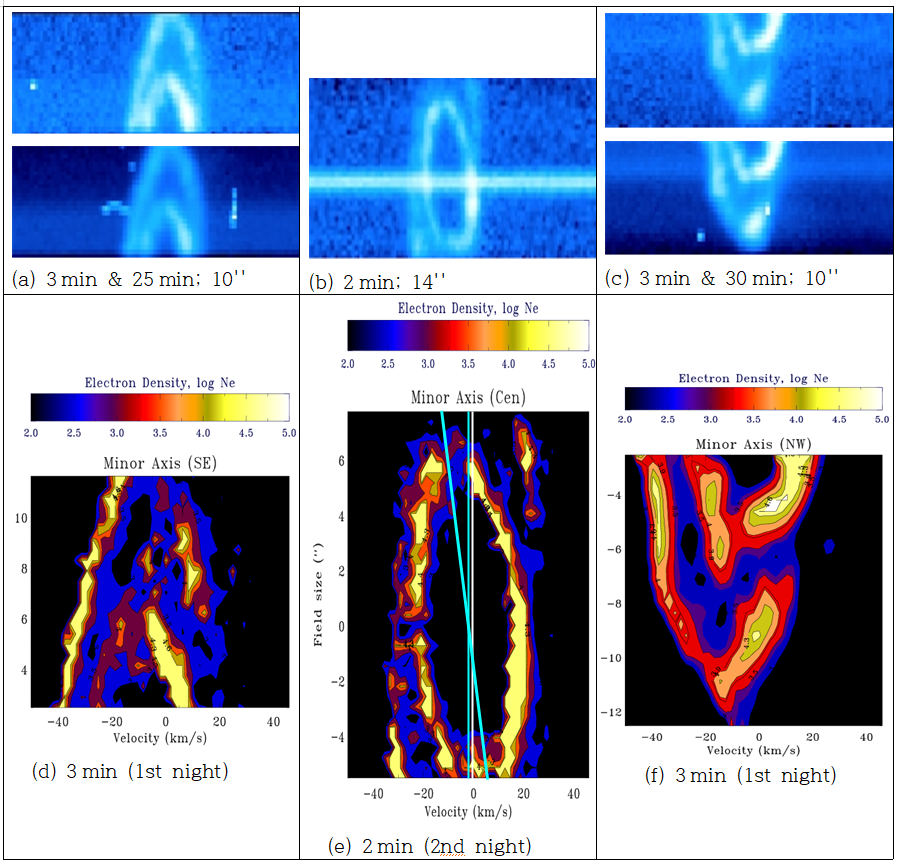}
\caption{Keck HIRES [{\sii}] 6731\AA\, spectral image and derived densities along the minor axis  {(PA = +347$^{\circ}$). }   
(a), (b), and (c): [{\sii}] 6731\AA\, logarithmic  scale flux of the S-SE rim, CSPN, and N-NW rim spectral images (see slit positions in Figure~\ref{fig1}). The slit lengths are 10$''$, 14$''$, and 10$''$  (2.6 pixels/1$''$), respectively. (d) -- (f): derived electron density P-V maps. Minimum flux cuts are  1.5\% for (d),  4\% for (e),  and  2\% for (f), respectively,   with respect to the absolute maximum fluxes. {False  dark colors indicate  
{regions with no  [{\sii}] flux  (that is, very noisy or below the level of flux minimum or log~$N_{\rm e}$ = {2.5} dex),} while the false-white color on the color bar indicates  values exceeding the maximum value ({4.6} dex) of the density (but no region exists exceeding this limit)} in the P-V maps.
The white vertical line is the position of the radial velocity of $-48.91$ {\kms}, determined by the flux median position.  The sky-blue vertical line is the newly estimated kinematic centroid (see Section~\ref{sec:twosh}). See the text and Table~\ref{tbl-3}.
 }
\label{fig6}
\end{figure*}

The measured signal-to-noise (S/N) of the four peaks of the line profiles  in Figure~\ref{fig5} are 32.5 and 27.7 (main rings) and 5.26 and 10.4 (outer rings), which correspond to errors of 3.1\% and 3.6\% (main rings) and 17.8\% and 9.6\% (outer rings),  respectively. 
The signal is proportional to the flux intensity, but the noise in the continuum near the line remains about the same in one spectral order: the S/N $\sim$ $\sqrt{t  \times flux}$ for an exposure-time $t$.
The S/N ratio of the spectral line segment of a roughly Gaussian line profile is higher at the center of the line profile (or ring) but becomes lower at the edge. 
Weak [{\sii}] flux regions are low S/N regions, which  are marked in black in  Figure~\ref{fig6}.
There are zones with a ratio of [{\sii}] 6716\AA\,/[{\sii}] 6731\AA\, below 0.45, which corresponds to a high electron density of log~$N_{\rm e} \simeq$ {4.5} dex (see Figure~\ref{fig3c}).

\begin{figure*}
\centering
\includegraphics[width=100mm]{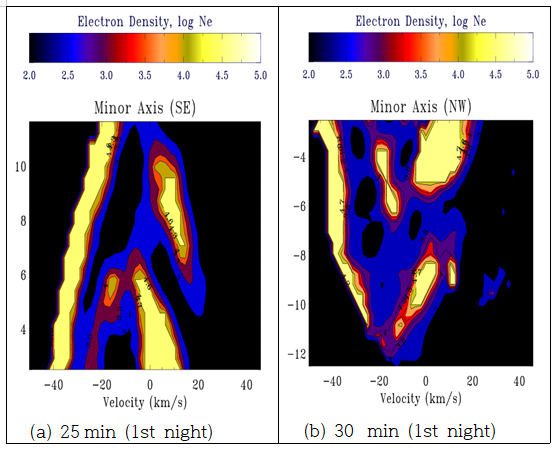}
\caption{Keck HIRES [{\sii}] 6731\AA\, spectral image and derived densities.  (a) 25 min and (b) 30 min. Minimum flux cuts  with respect to the maximum fluxes are 1.5\% for   (a)  and  2\% for  (b). 
 See  Figure~\ref{fig6} notes and the text.
 }
\label{fig7}
\end{figure*}

Figure~\ref{fig6} shows the [{\sii}] 6731\AA\, 2-D-spectral images in logarithmic flux scale  and P-V density distribution contour maps at the three slit positions: the S-SE rim, the CSPN, and the N-NW rim, secured with the slit entrances as in Figure~\ref{fig2} along the minor axis  (PA =  $+347^{\circ}$). Note that the present Keck 2-D spectral images correspond to a PA = +169$^{\circ}$ [{\nii}] image by \citet{sab04} (see their Figure~3), but with the top and bottom of the images reversed. The logarithmic scaled 6731\AA\, spectral images appear to be simple in shape, but the P-V density maps look very fuzzy and messy around the low S/N perimeter. The high S/N images in Figure~\ref{fig7} show two main shell structures well. However, the 25- and 30-minute exposure density images in Figure~\ref{fig7} seem to obscure detailed information about the main shell boundary.
The discordance may reflect the difficulty of getting an appropriate result due to the seeing conditions in the low S/N region and flux minimum cut levels. 
Hence we prefer Figure~\ref{fig6}. 
There is some overlap between the spectral image secured with the slit placed at the CSPN in Figure~\ref{fig6} (b) and those of  Figures~\ref{fig6} (a) and (c).

Figure~\ref{fig6} also shows the  P-V density maps of short-time exposures of 3, 2, and 3 minutes,  while 
Figure~\ref{fig7} shows maps from long-time exposures of 25 and 30 minutes, respectively.  
In Figures 6 and 7, we set the range of the color bar legend to Log~$N_{\rm e}$ = 2 -- 5 dex. The densities are mainly in a range of  Log~$N_{\rm e}$ = 2.8 -- 4.6 dex, except for Figure 6 (f) which shows one region of 4.7 dex. However, Figure 7 (b) shows that the densities of the same area are 4.6 dex, whose difference is within the marginal error of 0.2 dex. Considering all the circumstances into account, we conclude that the maximum density is about 4.6 dex.

There is a notable difference in density structure between Figures~\ref{fig6} (e) and (f), \ie the lower part of the two shells. 
It is unavoidable to see relatively large noise in short exposure two-dimensional images since we subtracted the almost-linear continuum level (i.e., the first-order spline). In other words, we only correct the continuum level, not the noise's details.

 The shorter the exposure time, the greater the noise level of the obtained image.   
From many trials and errors to avoid poor S/N zones in the maps, we set  the minimum flux level at 1.5\% level of the maximum flux for the S-SE maps in Figures~\ref{fig6} (d) and~\ref{fig7} (a);   slightly higher 2\% level for the N-NW maps in Figures~\ref{fig6} (f) and~\ref{fig7} (b); and  4\% level  for the center map in Figure~\ref{fig6} (e),  respectively.
The lower minimum flux level at 1.5\% level in Figure~\ref{fig6} (d) seemed to bring out some interesting features between the two shells more clearly, which are not discussed in this study. 
Different relative flux limits imply a more or less constant absolute flux limit to avoid low S/N { zones in the short spectra.}

Table~\ref{tbl-2} lists the  measured (mean) S/N values for the peaks of the [{\sii}]6716 and [{\sii}]6731 fluxes involving the density derivation. 
The designations (d) -- (f) in column (1), indicate the {panels}
in Figure~\ref{fig6}, and the arc-second values designate the distance from the center.  
Columns (2) -- (5) give the S/N ratios of  the  stronger [{\sii}] line profiles of the main shell and the outer shell  along the minor axis  at selected spatial positions. We distinguished the blue and red components of the main and outer   rings. 
The presence of the main and outer shells is evident in the spectral images. We only provide the S/N values for the most substantial peak of the shells whenever their presence is apparent from the images and the spectra, as seen in Figure~\ref{fig4}.
Note that the S/N ratio of the long exposure is much higher than that of the short exposures at 201 (25 min.) (not shown)  vs. 66.5 (3 min.) at the same $-4''$ position of the receding main shell. Measurement errors of the fluxes are similarly large (5 -- 10\%) for short exposures of 2 -- 3 minutes  in Figures~\ref{fig6} (d), (e), and (f);  and small (1 -- 5\%)  for long exposures of 25 and 30 minutes in Figures~\ref{fig7} (a) and (b). 
The difference between deep exposure and short exposure images is mainly in the peripheral boundary or the weak intensity region between the two main shells.

\begin{figure*}
\centering
\includegraphics[width=80mm]{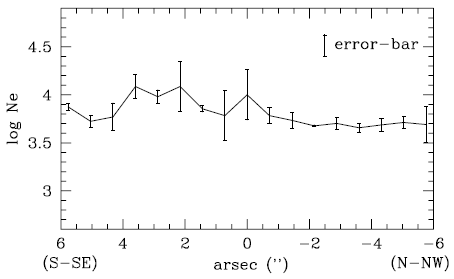}
\caption{A plot of density as a function of slit position. 
The x-axis indicates the slit position along the slit length (Figure~\ref{fig6} (e)), while the y-axis shows a mean density value from the slit position. The mean densities derived along the slit length are 3.69 to 4.09 ($\pm$0.14) {\cc}, which are much lower than the local density maxima as in Figure~\ref{fig6}.   See the text.   
 }
\label{fig8}
\end{figure*}

\begin{table}
\centering
\caption{S/N ratio at selected positions. Column (1): offset y position. See Figure~\ref{fig6}.
}
\begin{tabular}{@{}ccccc@{}}
\hline 
\hline
	 position              &  Main(blue) & Main(red)        & Outer(blue)   & Outer(red)     \\             
\hline
   &   S/N      &  S/N      & S/N       &  S/N        \\                     
   {\bf (d)~~ }  &          &        &        &         \\                         
10$''$    &   x        &  x        &  19.7  &   20.4   \\
8$''$      &   x        &  x        &  16.5  &   12.5   \\
6$''$      &   10.1  &  16.0  &  7.75     &   5.1      \\
4$''$      &   23.3  &  36.6  &  18.6  &   x         \\                                      
        
\hline
{\bf (e)~~ }  &            &           &           &             \\
6$''$      &   18.5  &  16.7  &  3.31     &   57.1      \\ 
4$''$      &   32.5  &  27.7  &  5.62     &   10.4   \\ 
2$''$      &   22.2 &  21.6  &  9.55     &   x         \\ 
-2$''$     &   17.4  &  56.8  &  15.3  &   x         \\    
\hline
{\bf (f)~~ }  &            &           &           &             \\
-4$''$     &   33.8  &  66.5  &  22.3  &   x         \\ 
-6$''$     &   15.8  &  96.3     &  17.4  &   x         \\
-8$''$     &   x        &  x        &  15.7  &   18.0   \\
-10$''$    &   x        &  x        &  x        &   28.0   \\

\hline

\end{tabular} 
\label{tbl-2}
\end{table}

Figure~\ref{fig8} shows a plot of mean density as a function of position along the slit length for the central slit position.
Note that the slit size is 14$^{''}$ while the number of data points are 38. Hence, we chose the slit position from every other data point to mimic the seeing slit size, \eg  (2), (4), ..., (36). The derived values at the boundaries are ignored due to the uncertainty. 
{The Keck observations show the density structure along the line of sight, resolved in velocity.  The P-V density maps in Figure~\ref{fig6} show high local densities and considerable variations along the line of sight.  Such a local density variation is only available with observations that resolve the structure along the line of sight in velocity. }

The densities from the integrated slit cross-section indicate a much  narrower density range  of 3.69 to 4.09 ($\pm$0.14) {\cc}. 
These values are similar to other densities integrated over the line of sight in section~\ref{sec:sight}. 
The error bars indicate the density disagreement between the value at each chosen slit position and the three-point smoothing mean value, \eg a smoothing mean value for the slit  position (2) is obtained by averaging three values for positions 1, 2, and 3. The actual errors might be smaller since the integrated fluxes are larger than the local fluxes (see the S/N ratios obtained from each point in Table~\ref{tbl-2}). 
A comparison of Figures~\ref{fig6} (e) and~\ref{fig8} seems to show that the high dispersion slit spectral images would be an excellent tool for finding locally varying densities along the line of sight.

The range of local densities of the {main} and outer shells in Figure~\ref{fig6} are log~$N_{\rm e}$ $\simeq$ {2.8} --  4.6 dex [{\cc}] (see also Figure~\ref{fig7}).
Table~\ref{tbl-3} lists the local density ranges of the two main shells. 
The last three lines of Table~\ref{tbl-3} give the density  derived from the integrated flux in Figure~\ref{fig4} and the Lick HES data (Hyung and Aller 1995a, b) using  
the Cloudy P-I code result.  The [{\sii}] 6716/6731\AA\, line ratios are given in brackets, indicating presumably high electron densities in the  [{\sii}] emission zones.

The S-SE rim (or shell) of Figure~\ref{fig6} (d) and the N-NW rim of Figure~\ref{fig6} (f) 
indicate the densities in both the {main} and outer shells,  which are as high as   log~$N_{\rm e}$ = {4.6} dex (40\,000 {\cc}). 
The inner {main}  shell consists of a complex substructure of $\sim$ $\log~N_{\rm e}$ $\simeq$ 2.8 -- 4.6 dex.

Our study shows that 
a wide variation of local densities from point to point along the line of sight in 
Figures~\ref{fig6} and~\ref{fig7}
is much more comprehensive than the integrated ones in Figure~\ref{fig8}. 
The approaching part of the outer shell shows a relatively thinner Arc shape with  densities of  $\log~N_{\rm e}$ $\simeq$  3.0 -- 4.6 dex 
(see Figures~\ref{fig6} (d) and (f)).  The approaching outer shell is well defined in both spectral images and P-V maps. 
The receding part of the outer shell shows a broken outer shape. In Figure~\ref{fig6} (b), the receding side of the main shell below the central star shows a well-defined bright primary form, which is also a denser region. The density range is about the same as the approaching part,  $\log~N_{\rm e}$ $\simeq$ 2.8 -- 4.6 (or 4.7) dex.    

The HST [{\nii}] image in Figure~\ref{fig1} shows many bright small-scale  features, all of which must be due to density enhancements in the optically thin NGC 7009. 
If the line of sight is unusually long, for example, a considerably long path length occurs at the edge of the spherical shell, we also expect a brightness enhancement (\ie the bright rim). 
Apparently, the relatively fainter outer shell does not form a whole circle in the P-V density maps, especially in Figure~\ref{fig6} (f). 
Figures~\ref{fig6} (a)--(c) show that the receding part of the outer shell is relatively faint. The P-V density maps show that the receding part of the outer shell is not a perfect arc (see the weak flux zone of the spectral images Figures~\ref{fig6} (a) and (c)  and  the S-SE and N-NW P-V density maps in Figures~\ref{fig6} (d) and (f)).

According to \citet{sab04,sab06}, NGC 7009 is an optically thin PN that is powered by a luminous post-AGB star at a high temperature. The outer shell gas along the minor axis might have a change in the ionization state of the gas, as occurs naturally due to the greater distance of the outer shell from the source of ionizing radiation. Although S$^+$ is a minority ion throughout the main shell, it may be more abundant in the outer shell, perhaps the dominant ionization phase there.

\begin{table}
\centering
\caption{Electron densities of the equatorial shells along the minor axis. The density ranges of the shells and specific positions are given based on the P-V maps of Figure~\ref{fig6}. 
We list density error (dex) in parentheses, assuming about 5\% measurement error, at a peak flux position of  high S/N (HIRES).  $^*$: The density derived in the present result is not different from the earlier HES result given within the square parentheses (Hyung and Aller 1995b).   
See the text.  }

\begin{tabular}{@{}cccc@{}}
\hline 
\hline
Position & Log~$N_{\rm e}$ [{\cc}] &  P-V map  & Note     \\
\hline
   & $<{\rm {main}~shell}>$   &     &     \\
  whole   & 2.8($\pm$0.08) -- 4.6($\pm$0.20) &  Figure~\ref{fig6} &   \\
 &  $<{\rm outer~shell}>$ &          &     \\
      whole   & 2.8($\pm$0.08) -- 4.6($\pm$0.20) &    Figure~\ref{fig6} &    \\
 &  $<$ {integration }$>$ &          &     \\
5--6$''$   & 3.55($\pm$0.15)$^*$ & HES & w/ Figure~\ref{fig3c}   \\
  & [3.55($\pm$0.15), 4.3($\pm$0.20)]    & HES   &  \\ 
 4$''$ & 3.8($\pm$0.10) ($10^{2.8}$ -- $10^{4.5}$)  & HIRES 
  &   Figure~\ref{fig4}\\
\hline
\end{tabular} 
\label{tbl-3}
\end{table}

\subsection{Revised Expansion Velocities of Two Shells}  \label{sec:twosh}

\citet{sab04} constructed a 3-D morpho-kinematical model structure based on detailed P-V diagrams and Hubble Space Telescope imagery. They assumed a general law of the expansion velocity, V$_{\rm exp}$ = c$_1$ $\times R$ + c$_2$, where  c$_1$ is the acceleration expansion slope;  c$_2$ is the velocity at the center point when the velocity slope is extrapolated;  and $r$ is a spatial distance from the center to the line emission zone which may be found from an observed  distance of $r$ projected onto the sky.  
The radial velocity variations of the main and outer shells by \citet{sab06} show a linear relation where V$_{\rm exp}$ = $4.0 \times r''$ and $3.15 \times r''$, respectively, which are extrapolated to the central star (c$_2 \sim 0$, see their Figure 5). \citet{hyu14} revised V$_{\rm exp}$ by \citet{sab06} with the various Keck HIRES spectral images and the HST images which show
 a non-zero c$_2$ component  for the main shell.
 
\citet{ste09} also constructed a 3-D morpho-kinematical model structure 
using the data by  \citet{sab04, sab06}. Their study generally confirmed the results of \citet{sab06}, but they found evidence of non-homologous expansion velocities, \ie a  10 -- 15\% deviation  from a homogeneous radial expansion at the equatorial zone of the main shell.

The P-V kinematic map in Figure~\ref{fig6} shows that the outer shell is  distorted, as is the inner shell. \citet{sab04} argue that the position angle to get symmetric profiles is near 130 degrees.
As mentioned in previous studies such as \citet{sab04} or \citet{hyu14} (see their Figure~8), the expansion velocity variation of sub-shells over distance was determined for the two main shells. 
\citet{hyu14} incorrectly calculated the radius and velocity of the outer shell based on the appearance of the inner bright shell appearance. The kinematic center must be correctly re-estimated considering the outer shell's overall shape as well. 
In Figure~\ref{fig6} (e), we added two (white and sky-blue) vertical lines and one slightly slanted (sky-blue) line.
The white vertical line, which indicates the position of the radial velocity of $-48.91$  {\kms}, does not appear to be at the center of the two shells. Meanwhile, the newly estimated kinematic centroid (slanted blue-sky line) which is derived by the general appearance of both ellipsoidal shapes looks fine.
The slanted sky-blue line which divides the inner shell velocity ellipse corresponds to the median of the line profile shape (\ie the kinematical center).  As mentioned, the former radial velocity was derived based on the flux median of strong lines rather than the line shapes. The shape of the outer shell does not form a perfect ellipse. Still, the newly determined radial velocity, mainly by the main shell appearance, appears to be  suitable for the outer shell.

\begin{table}
\small
\caption{Expansion velocities vs. radial 
distance.
Radius units in columns (3) and (5): arcsec.
Velocity unit in columns (4), (6)-(7):{~\kms}.  The errors in $V_{exp}$ are less than 1.5{~\kms}. 
S04: for the main shell by \citet{sab04}.   
*{\hi}: $V_{exp}$ for the whole PN. See \citep{hyu14} and the text.} 
\vspace{-0.4cm}
\begin{tabular}{cccc ccc}\\
\tableline
Ion         & IP(eV) & R(main)  & V$_{\rm exp}$(main) & R(out)   & V$_{\rm exp}$(out) & S04        \\
\hline
{\heii}      & 54.4   &  4.00         & 19.80       &        &    -           &   18.3   \\
{\niii}      & 47.4   &  4.32   & 19.34        &        &    -           &   20.0    \\
$[${\ariv}$]$   & 40.7   &  4.62    & 19.19     & 10.99   &   24.34        &              \\
$[${\cliv}$]$   & 39.6   &  4.67      & 19.86     & 11.02   &   25.36        &         \\
$[${\oiii}$]$   & 35.1   &  4.90      &   19.77     &         &    -           &   20.4   \\
$[${\ariii}$]$  & 27.6   &  5.22       &   19.86     & 11.42    &   25.88        &   20.4     \\
{\he1}      & 24.6   &  5.36      &   22.80        &         &        &   20.0     \\
$[${\siii}$]$   & 23.3   &  5.41        &   19.98     & 11.56    &   26.50              &     \\
$[${\nii}$]$    & 14.5   &  5.81       &   20.67     & 11.85   &   27.60        &   20.4      \\
{\hi}       & 13.6   &   -           &   20.09*       &         &    -          &   20.8     \\
$[${\sii}$]$    & 10.4   &   6.00     &   21.70    & 12.00   &   30.00        &   21.0       \\
\tableline
\end{tabular}
\label{tbl-4}
\end{table}

\citet{hyu14}'s values should be re-evaluated based on the latter centroid.  Table~\ref{tbl-4} lists the expansion velocity versus radius obtained by recalculating the radius shown in Table 4 of \citet{hyu14}. 
The present kinematic centroid implies that  the {main} [{\sii}] sub-shell expands with V$_{\rm exp}$([{\sii}]) = 21.7 {\kms}, indicating that \citet{hyu14}'s expansion velocity (26.52 {\kms}) is overestimated  by $\triangle$V$_{\rm exp}$([{\sii}]) = 4.82 {\kms}.
Meanwhile, the outer [{\sii}] sub-shell expands with  V$_{\rm exp}$([{\sii}]) = 30.0 {\kms}, less than \citet{hyu14}'s estimate (42.14 {\kms}) by  $\triangle$V$_{\rm exp}$([{\sii}]) = 12.14 {\kms}. 
Thus, we applied the velocity difference of 
$\triangle {\rm V}_{\rm exp}$(M) = $-$4.82 {\kms} to the expansion velocities of the stratified main sub-shells of the {main} shell and 
similarly, a difference of $\triangle$V$_{\rm exp}$(O) = $-$12.14 {\kms} to those of the outer shell. 
We re-estimated the radii based on the new kinematic center: the main shell in Figure~\ref{fig6} (e) implies that the radius of the main [{\sii}] sub-shell is 6$''$, the same as in \citet{hyu14}, 
whereas the radius of the outer [{\sii}] sub-shell, determined by Figures~\ref{fig6} (d) and (f), decreased  by 1.00$''$ from 13.0$''$ to 12.0$''$. 
The radius of the outer [{\sii}] sub-shell and all other sub-shells should be changed by $\triangle r$ = -1.0$''$, as shown in column (5).
The current  estimate of the outermost radius of the outer shell is close to 11.8$''$ by \citet{sab04}. 
Here, `M' and `O' represent the main and outer shells. 
Note that the derived expansion velocities agree better with those of \citet{sab04}.

\begin{figure}
\centering
\includegraphics[width=70mm]{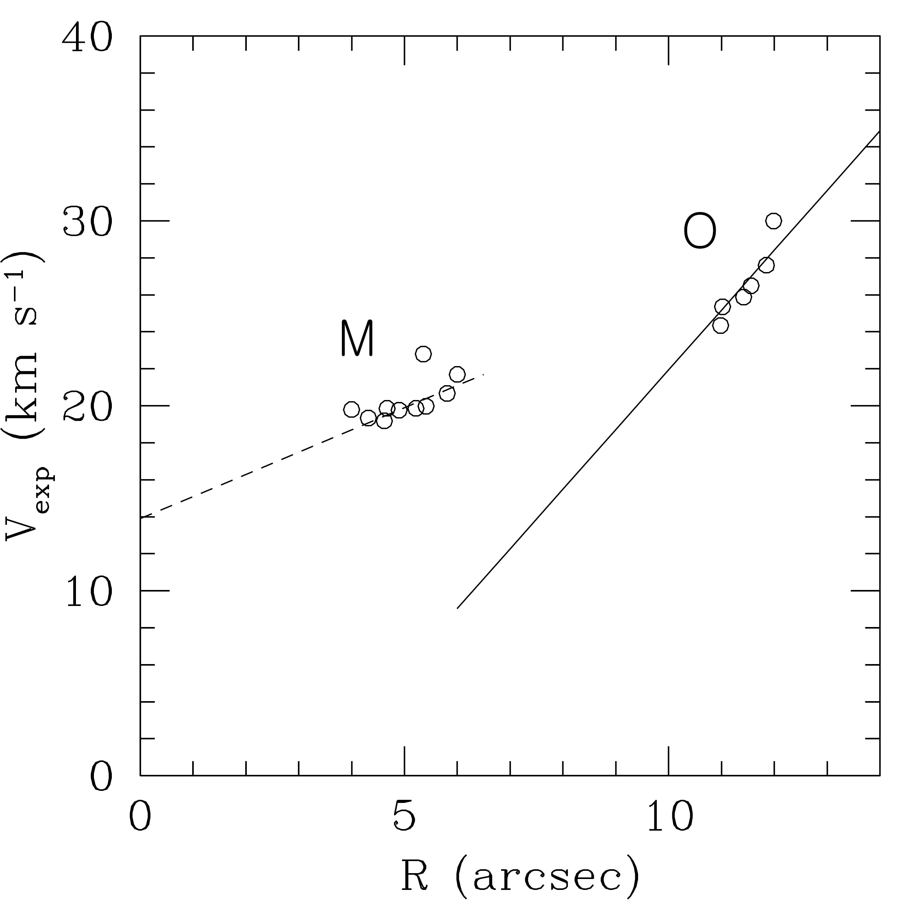}
\caption{ The Hubble-type expansion velocities of the main (M) and outer (O) shells along the minor axis with two trend lines, refined from \citet{hyu14}. When drawing trend and regression lines, He and H data were excluded. See Table~\ref{tbl-4} and the text.  
 }
\label{fig9}
\end{figure}

Figure~\ref{fig9} shows the velocity data points and two linear regression fits excluding the {\he1} and {\heii} lines that best represent the scattering of the expansion velocities of the sub-shells responsible for each ion. 
Using the linear least square function (\eg MS Excel’s LINEST: V = c$_1$ $\times$~r$''$ + c$_2$), the 
fitting yields (i)   to the main shell data: 
V$_{\rm exp}$(M) = 1.20($\pm$0.25) $\times$ r$''$ + 13.90($\pm$1.28) {\kms} and (ii)  to the outer shell data:  V$_{\rm exp}$(O) = 3.19($\pm$0.56) $\times$ r$''$  $-$10.36($\pm$6.37) {\kms},  where the value in parenthesis is 1 $\sigma$ (see Figure~\ref{fig9}). 
When finding the trend-lines (or slopes), we excluded the He and H line velocities representing large areas.
The [{\sii}] has such a larger deviation than other lines. We also excluded the [{\sii}] line for the outer shell.  With the [{\sii}] included, the fitting gives a very unusual trend line, 
V$_{\rm exp}$(O) =  4.45($\pm$0.90)  $\times$ r$''$  $-$24.41($\pm$10.28) {\kms}.  

In a post-AGB evolutionary stage, it is known that the radiative pressure of the central star pushes the ejected shell, so the shell reaches a terminal velocity (V$_t$) at a distance of 10 times the AGB radius  (\citealt{1981K,2000K, 2001F}).
The linearity means that the acceleration mechanism 
seems to have worked after leaving the central star. 
However, the c$_2$ values at the center are non-zero in both the inner main and outer shells. 
The main shell shows c$_2$ = 13.90 $\pm$1.28 {\kms} ($>$ 0).  Such a positive velocity at the center indicates the terminal velocity occurring in a post-AGB envelope in the early time or evidence of acceleration due to the hot bubble in the inner zone.

The intersection at the central position is related to the terminal velocity of post-AGB. In contrast, the slope of the trend line is associated with the acceleration (+) or deceleration (-) of the expanding sub-shells.
The slope of the outer  shell expansion velocity intersects the zero velocity, V$_{\rm exp}$(O) $\sim$ 0 (or up to +2.38) {\kms} within 2 $\sigma$ = 12.74 {\kms}.
A simple positive terminal velocity and following continuous accelerations after post-AGB cannot accommodate such a negative velocity.

There is very little difference between the results here and those in \citet{hyu14}: Both studies show that the main shell had a positive velocity intercept and the slopes of the relations are similar. The outer shell in the two studies had a velocity intercept compatible with zero at the center within 2$\sigma$, and the slopes show almost the same tendency.
The radius of the outer shell and, as a result, derived velocities are more similar to those of \citet{sab04}.

The 2D P-V density map from the HIRES spectral images along the major axis (Paper II, to be submitted) and the analysis by \citet{ste09} and \citet{sab04} show the presence of two torus shells at low latitudes near the equator, which are larger and faster than the presently studied equatorial outer shell. 
Both Sabbadin (2004) and the present study show that the outer shell along a minor axis has a positive slope, \ie a sign of acceleration.
Perhaps, the fast stellar wind caused an acceleration of the main shell, but not the outer shell at the equator, \ie deceleration after acceleration.

\citet{2001hyung} presented the case where NGC 6543 accelerated after reaching the terminal velocity after post-AGB.
One likely interpretation is that the gradually increasing slope made it look as if the terminal velocity was negative. 
When the utmost outer shell was detached from the earlier post-AGB star, the sub-shells expanded at a relatively small (\ie nearly zero) velocity, and as a result, the trend line met a very small positive value near the center at that time.  Later, when the second (now Main) shell emerged from a hot central star, the former (now Outer) shell experienced a relatively large  acceleration (or it became a large slope of c$_1$) with the current weird acceleration pattern.

The expansion velocities of [{\sii}] and other lines in
Figure~\ref{fig9} are from the stratified sub-shells. Although they generally show Hubble-type acceleration in both the  main and the outer shells, the kinematics of multiple shells is not included in the `interacting-winds' theory by \citet{1978kwok}.

\section{Conclusions \label{sec:con}}

Using the [{\sii}] 6716/6731\AA\, and  fixing the electron temperature, $T_{\rm e}$= 10\,000 K, we derived densities of NGC 7009 along the minor axis with the Cloudy code by \citet{2017RM}.  
From the derived P-V maps of [{\sii}] density, we were able to clearly distinguish two shells 
and find their densities and expansion velocities. 
Along the minor axis, the local density range of the {main} and the outer shells is log~$N_{\rm e}$ $\sim$ 2.8 -- {4.6} dex. However, Figure~\ref{fig6} (f) shows at least one region of log~$N_{\rm e}$ $\sim$ 4.7 dex in the inner main shell. 
The local densities derived from the two-dimensional P-V map decomposed by radial velocities along the line of sight show a more considerable variation than those obtained from the integrated fluxes along {the line of sight.  }

The [{\sii}] 6716/6731\AA\, ratio of the integrated spectrum 
for one pencil-beam at the position of 4$''$ indicates log~$N_{\rm e}$ $\sim$ 3.8 dex in Table~\ref{tbl-3}, whereas the 2-D spectral image analysis shows that the density values have a range of log~$N_{\rm e}$ $\sim 2.8$ -- {4.5} dex for the same position, confirming that the actual density fluctuation is significant in
two main shells. 
While NGC 7009 is not a high-density PN, there appear to be local regions where the [{\sii}] lines indicate high densities.

Since the density of NGC 7009 shown by other lines is not all high, our results likely indicate that the higher density occurs at the outer boundary of the shell region rather than the entire PN shell. 
Figures~\ref{fig6} (a) and (c) show this very clearly, as there is [{\sii}] emission from the volume between the two shells.
\citet{sab04} presented P-V maps for the [{\nii}] line which also show a double-shell structure at the minor axis. 

As mentioned earlier, NGC 7009 at PA = 169$^{\circ}$ (apparent minor axis) is an excellent example of an optically thin PN powered by a luminous post-AGB star at a high temperature \citep{sab06}.
{The inner shell morphology } may form due to the mass contrast, while the outer shell may form due to a change in ionization state. 
In any case, [{\sii}] is a useful tracer of the density wherever found in NGC 7009 or other PNe. 
Although the  spectral images of the [{\sii}] lines do not represent the whole system, they would provide many clues to the kinematics of the PN. 
Based on the new kinematic center, we derived the kinematics of the two shells.

In this study, we drew two basic conclusions. First, the local density has an extensive range along the line of sight 
Second, a velocity jump exists between the two shells, and different kinematic characteristics appear in their acceleration mechanisms. Apparently, the hot bubbles accelerated the main shell but had little effect on the outer part, especially the outer shell along the minor axis.
The well-known Hubble-type acceleration mechanism explains the expansion velocities of the main shell. 
However, the outer shell is not accommodated by such simple accelerations, suggesting that the physical state of the central star must have been very different in the past period. 
Future studies also need to investigate how the slit spectral images of other low excitation lines will differ from the densities obtained on the integrated lines.

\acknowledgments

S.H. is grateful to the late Prof. Lawrence H. Aller of UCLA, who carried out the Keck HIRES observation program together.  S.H. also would like to thank F. Sabbadin for encouraging by sending a paper by \citet{sab04} in 2004 and  Prof. Francis P. Keenan at the Queen's University of Belfast for providing additional information on the [{\sii}] atomic data.
We express our heartfelt gratitude to the anonymous referee for reviewing our paper. We thank Dr. Sung, E.-C. at KASI and Dr. Son, D.-H. at SNU for their help with the 2-D spectral data reduction.
S.-J.L. and S.H. would like to acknowledge support from the Basic Science Research Program through the National Research Foundation of Korea (NRF 2020R1I1A3063742; NRF 2017R1D1A3B03029309).
M.O. was supported by JSPS Grants-in-Aid for Scientific Research(C) (JP19K03914).

\vspace{5mm}

\facilities{Keck:10m, HDS}

\software{IRAF (Tody 1986,1993), Starlink \citep{2014star},            Cloudy \citep{2017RM}}

%\bibliography{n7009}{}

\bibliographystyle{aasjournal}

\end{document}